**Numerical Representation of Preferences over Random Availability Functions: A Note**

**By**

**Somdeb Lahiri**

**ORCID: https://orcid.org/0000-0002-5247-3497**

**(Formerly) PD Energy University (EU-G)**

**(somdeb.lahiri@gmail.com)**

**April 26, 2025.**

**This version: May 3, 2025.**

**Abstract**

We interpret a fuzzy set as a random availability function and provide sufficient conditions under which a preference relation over the set of all random availability functions can be represented by a utility function.

**1. Introduction:** The third paragraph of section 2 of Basu (1984) begins with the statement "It is important to realize that given a fuzzy set A, A(x) is not the probability that x belongs to A'. Keeping this cautionary statement in mind, we interpret a fuzzy set as a "Random Availability Function." A random availability function assigns to each alternative the probability with which it will be available. While a lot of work pertaining to applications of fuzzy set theory to the social and decision sciences exists, we have chosen Basu (1984) as the sole reference for it for two reasons. The first reason is that to the best of our knowledge, Basu (1984) is the first one to apply fuzzy set theory to economics. The second reason is, that except for the definition of a random availability function, there is nothing in what follows that has anything to do with fuzzy set theory. This paper is an attempted contribution to decision theory and therefore to applied probability theory in a tenuous sense, however insignificant the contribution may be. None of the concepts discussed here are related to stochastic choice functions, discussed in Barbera and Pattanaik (1986), Border (2007), Falmagne (1978), Fudenberg, Iijima and Strzalecki (2015) and Mas Colell (2007), followed by many others.

Consider a restaurant whose menu includes fish, chicken and lamb, though it is rarely the case that all three are available on the same day. "Aaaj nahi bunaa/buni hain" ((Its) not been made today) is a pretty common response one gets at restaurants in Ahmedabad, particularly if you are looking for your favourite dish. Clearly, before choosing to go the restaurant, one has to conjecture the probabilities with which each item would be available that day and hence the problem concerns choosing or not choosing a random availability function. A very common answer one gets at our neighbourhood "mithai" shop that is supposed to be selling "malai peda" made with stevia (instead of sugar) is "aaj nahin hai lekin kal zaroor mil jaayega" ((Its) not there today, but it will certainly be available tomorrow). This, is yet another example of a random availability function, where sugar-free malai peda being available the same day and sugar-free malai peda being available at a future date are really two different items, with the probability of getting the first item being zero and the probability of getting the second item being positive. The reason for such unwarranted unavailability is rooted in a culture that prioritizes "trading" over the "quality of the product/service that is being traded". However, the fact remains that both are real world examples of random availability functions. The natural decision-theoretic problem in this context is choosing one or more random availability functions from a given non-empty set of random availability functions. Unfortunately, the concept of a random availability function of the kind we discuss here, does not seem to have been discussed so far in the available literature on decision theory, either at a conceptual level or at a rigorously theoretical level. Gilboa (2011) provides a very lucid discussion of concepts in decision theory and a considerably more rigorous discussion of the relevant theory in available in Gilboa (2009).

In order to choose from a non-empty set of random availability functions, the perspective adopted in decision theory would require a "preference relation" over the set of random availability functions that one could choose from, so that the chosen functions are perceived to be "at least as good" as all functions that are available for choice. This process gets considerably simplified, if the preference relation has a numerical representation, so that the problem of choice reduces to "constrained maximization" of a real-valued function.

In this note, we provide sufficient conditions for a preference relation on the set of random availability functions to allow a numerical representation. The two sufficient conditions we invoke for our main result are weak dominance and weak continuity. One random availability function "strongly dominates" another, if the probability of availability of each alternative in the former is strictly greater than the corresponding probability in the latter. Preferences are said to be "weakly dominant" if whenever one random availability function strongly dominates another random availability function, then the first is strictly preferred to the second. Weakly dominant seems to be harmless and a perfectly plausible assumption. Weak continuity is a technical property, "stronger" variants of which are of regular use in economics and hence in addition to being quite reasonable, it is unobjectionable as well. After having proved lemma 1, the proof of the main result is almost identical to the proof of the result concerning numerical representation of preferences in the context of consumer demand theory that is available in Varian (1978). Hence, the originality of our result is entirely in the concept of a random availability function that it is relevant for and hardly any in the underlying mathematics used for its proof.

**2. The Model:** Our notations largely follow those in Basu (1984).

Let $\mathbb{N}$ denote the set of natural numbers, $\mathbb{R}$ the set of real numbers and $\mathbb{R}_+$ the set of non-negative real numbers. Let $[0, 1] = \{\alpha \in \mathbb{R} | \ 0 \leq \alpha \leq 1\}$ and $(0, 1) = \{\alpha \in \mathbb{R} | \ 0 < \alpha < 1\}$.

Let X be a non-empty finite set of alternatives containing at least two alternatives.

A "**random availability function" (RAF)** (on X) is a function A: X→ [0, 1], such that for each x∈X, A(x) is the probability that x is available and 1-A(x) is the probability that x is not available.

Let $\mathcal{P}$(X) denote the set of all RAF's on X. Thus $\mathcal{P}(X) = [0, 1]^X$.

Let $\succcurlyeq$ be binary relation on $\mathcal{P}(X)$. Instead of writing $(A, B) \in \succcurlyeq$, where $A, B \in \mathcal{P}(X)$, to economize on notation we will write $A \succcurlyeq B$.

We assume that $\succcurlyeq$ is reflexive (i.e., $A \succcurlyeq A$ for all $A \in \mathcal{P}(X)$), connected (i.e., for all $A, B \in \mathcal{P}(X)$, either $A \succcurlyeq B$ or $B \succcurlyeq A$) and transitive (i.e., for all $A, B, C \in \mathcal{P}(X)$: $[A \succcurlyeq B$ and $B \succcurlyeq C]$ implies $[A \succcurlyeq C]$.

Such a binary relation $\succcurlyeq$ is called a **preference relation** with the interpretation that if for $A, B \in \mathcal{P}(X)$, it is the case that $A \succcurlyeq B$, then A is "**at least as good as**" B.

The asymmetric part of $\succcurlyeq$, denoted by $\succ$ is interpreted as "**is strictly preferred to**" and the symmetric part of $\succcurlyeq$, denoted by $\sim$ is interpreted as "**no different from**".

Given RAF's A and B, A is said to **strictly dominate** B, if $A(x) > B(x)$ for all $x \in X$.

$\succcurlyeq$ is said to be "**weakly dominant**" if for $A, B \in \mathcal{P}(X)$: [A strictly dominates B] implies $[A \succ B]$.

Let $A^*$ be the RAF such that $A(x) = 1$ for all $x \in X$ and let $A^0$ be the RAF such that $A^0(x) = 0$.

If $\succcurlyeq$ is weakly dominant then $A^* \succ A^0$.

For all $t \in (0, 1)$, let $tA^*$ be the RAF such that $(tA^*)(x) = tA^*(x)$ for all $x \in X$.

It is easy to see that $\mathcal{P}(X)$ is a "**closed set**" in the sense that if $\langle A^{(n)} | n \in \mathbb{N} \rangle$ is a sequence in $\mathcal{P}(X)$ converging to A (i.e., $A: X \rightarrow \mathbb{R}$ such that for all $x \in X$, $\lim_{n \to \infty} A^{(n)}(x)$), then $A \in \mathcal{P}(X)$.

$\succcurlyeq$ is said to be **weakly continuous** if for all sequences $\langle A^{(n)} | n \in \mathbb{N} \rangle$ and $\langle B^{(n)} | n \in \mathbb{N} \rangle$ in $\mathcal{P}(X)$ converging to A and B respectively, then $[A^{(n)} \succ B^{(n)}$ for all $n \in \mathbb{N}]$ implies $[A \succcurlyeq B]$.

There is a stronger version of continuity that is normally used in decision theory, that is obtained by replacing the strict preference relation with the "at least as good as" relation in the definition of weakly continuous preferences. However, this stronger version is not required to obtain our main result.

A function $u: \mathcal{P}(X) \rightarrow \mathbb{R}$ is said to be a **utility function for (numerical representation of)** a preference relation $\succcurlyeq$ if for all $A, B \in \mathcal{P}(X)$: $[A \succcurlyeq B]$ <u>if and only if</u> $[u(A) \geq u(B)]$.

**Example:** Let $\pi: X \rightarrow \mathbb{R}_+$ be a function such that for all $x \in X$, $\pi(x)$ is the pay-off from choosing x. Thus, if $x \in X$ is the desired alternative, then the expected pay-off from A is $\pi(x)A(x)$. Let $u: \mathcal{P}(X) \rightarrow \mathbb{R}$ be such that for all $A \in \mathcal{P}(X)$, $u(A) = \max_{x \in X} \pi(x)A(x)$ and $\succcurlyeq$ be the preference relation on $\mathcal{P}(X)$ such that for all $A, B \in \mathcal{P}(X)$: $A \succcurlyeq B$ if and only if $u(A) \geq u(B)$. Clearly, u is a utility function for $\succcurlyeq$.

**3. The Main Result:**

We first prove an important lemma.

**Lemma 1:** Let $A, B \in \mathcal{P}(X)$ and suppose $A(x) \geq B(x)$ for all $x \in X$. If $\succcurlyeq$ is weakly dominant and weakly continuous, then $A \succcurlyeq B$.

**Proof:** If $A = B$, then the reflexivity of $\succcurlyeq$ implies the lemma. Hence suppose $A \neq B$.

Let $I_1 = \{x\in X|\ A(x) = 1\}$, $I_2 = \{x\in X|\ B(x) = 0\}$ and $I_3 = \{x\in X|\ A(x) < 1, B(x) > 0\}$.

Clearly for all $x\in I_3$, it must be the case that $0 < B(x) \leq A(x) < 1$.

Let $\varepsilon \in(0, 1)$ be such that $B(x) - \varepsilon > 0$ for all $x\in I_3$, in the event that $I_3 \neq \phi$.

For $x\in X$ satisfying $B(x) < A(x)$, let $B^{(n)}(x) = B(x)$ and $A^{(n)}(x) = A(x)$ for all $n\in \mathbb{N}$.

For $x\in X$ satisfying $B(x) = A(x)$, let $B^{(n)}(x) = B(x) - \frac{1}{2n}$, $A^{(n)}(x) = A(x)$ for all $n\in \mathbb{N}$ if $x\in I_1$, let $B^{(n)}(x) = B(x)$, $A^{(n)}(x) = A(x) + \frac{1}{2n}$ for all $n\in \mathbb{N}$ if $x\in I_2$ and let $B^{(n)}(x) = B(x) - \frac{\varepsilon}{2n}$, $A^{(n)}(x) = A(x)$ for all $n\in \mathbb{N}$ if $x\in I_3$.

Clearly the sequences $<A^{(n)}|n\in\mathbb{N}>$ and $<B^{(n)}|n\in\mathbb{N}>$ are in $\mathcal{P}(X)$ and converge to A and B respectively. Further, $A^{(n)}(x) > B^{(n)}(x)$ for all $x\in X$ and $n\in\mathbb{N}$.

Since $\succcurlyeq$ is weakly dominant, $A^{(n)} \succ B^{(n)}$ for all $n\in\mathbb{N}$.

Thus, since $\succcurlyeq$ is weakly continuous, it must be the case that $A \succcurlyeq B$. Q.E.D.

Now we proceed to our main result which and its proof are very similar to the usual result on numerical representation of consumer preferences by Wold (1943) that is used in demand theory and its proof that is available as a proposition on page 82 of Varian (1978).

**Proposition 1:** Suppose $\succcurlyeq$ is weakly dominant and weakly continuous. Then $\succcurlyeq$ has a utility function.

**Proof:** Let $A\in\mathcal{P}(X)$ and let $\mathcal{U}(A) = \{t\in[0,1]|tA^* \succcurlyeq A\}$.

Since $A^* \geq A \geq A^0$, by lemma 1, $1\in\mathcal{U}(A)$ and 0 is a lower-bound of $\mathcal{U}(A)$.

By the completeness axiom of the real number system, $\mathcal{U}(A)$ has a greatest lower bound which we denote by u(A). Clearly, $u(A) \in[0,1]$.

We want to show that $u(A)A^*\sim A$.

Towards a contradiction suppose it is ***not the case that*** $u(A)A^*\sim A$.

First suppose, $u(A)A^*\succ A$. Thus, it must be the case that $u(A) > 0$, since by lemma 1, $A \succcurlyeq A^0$. Let $\varepsilon\in(0, 1)$ be such that $u(A) - \varepsilon > 0$.

Clearly, $(u(A) - \frac{\varepsilon}{n})A^*\in\mathcal{P}(X)$ for all $n\in\mathbb{N}$ and the sequence $<(u(A) - \frac{\varepsilon}{n})A^*|\ n\in\mathbb{N}>$ converges to $u(A)A^*$.

If $A \succ (u(A) - \frac{\varepsilon}{n})A^*$ for all $n\in \mathbb{N}$, then applying weak continuity to the constant sequence all whose terms are A and the sequence $<(u(A) - \frac{\varepsilon}{n})A^*|\ n\in\mathbb{N}>$, we would get $A \succcurlyeq u(A)A^*$, contradicting our assumption that $u(A)A^*\succ A$.

Thus, it must be the case that for some $n\in \mathbb{N}$, $(u(A) - \frac{\varepsilon}{n})A^* \succcurlyeq A$ and hence $u(A) - \frac{\varepsilon}{n} \in\mathcal{U}(A)$. Since $u(A) - \frac{\varepsilon}{n} < u(A)$, such a result would contradict u(A) is a lower bound of $\mathcal{U}(A)$.

Hence, $u(A)A^*\succ A$ is not possible.

Now suppose $A \succ u(A)A^*$. Thus, $u(A) \notin \mathcal{U}(A)$ and $u(A) < 1$, since by lemma 1 $A^* \succcurlyeq A$. Let $\varepsilon \in (0, 1)$ be such that $u(A) + \varepsilon < 1$.

Clearly, $(u(A) + \frac{\varepsilon}{n})A^* \in \mathcal{P}(X)$ for all $n \in \mathbb{N}$ and the sequence $<(u(A) + \frac{\varepsilon}{n})A^* | \, n \in \mathbb{N}>$ converges to $u(A)A^*$.

If $(u(A) + \frac{\varepsilon}{n})A^* \succ A$ for all $n \in \mathbb{N}$, then applying weak continuity to the constant sequence all whose terms are A and the sequence $<(u(A) + \frac{\varepsilon}{n})A^* | \, n \in \mathbb{N}>$, we would get $u(A)A^* \succcurlyeq A$, contradicting our assumption that $A \succ u(A)A^*$.

Thus, it must be the case that for some $n \in \mathbb{N}$, $A \succcurlyeq (u(A) + \frac{\varepsilon}{n})A^*$.

Since, $t \in \mathcal{U}(A)$ implies $tA^* \succcurlyeq A$, by transitivity of $\succcurlyeq$ we get $tA^* \succcurlyeq (u(A) + \frac{\varepsilon}{n})A^*$ for all $t \in \mathcal{U}(A)$.

If $0 \leq t < u(A) + \frac{\varepsilon}{n}$, then $(u(A) + \frac{\varepsilon}{n})A^*$ strongly dominates $tA^*$ and since $\succcurlyeq$ is weakly dominant we get $(u(A) + \frac{\varepsilon}{n})A^* \succ tA^*$.

Thus, it must be the case that for all $t \in \mathcal{U}(A)$, $t \geq u(A) + \frac{\varepsilon}{n}$.

Thus, $u(A) + \frac{\varepsilon}{n}$ is a lower bound of $\mathcal{U}(A)$ and since $u(A) + \frac{\varepsilon}{n} > u(A)$, the definition of u(A) as the greatest lower bound of $\mathcal{U}(A)$ is contradicted.

Thus, it is not possible that $A \succ u(A)A^*$.

Since, $u(A)A^* \succ A$ is not possible, the fact that $\succcurlyeq$ is reflexive and connected implies that $u(A)A^* \sim A$.

Let $A, B \in \mathcal{P}(X)$. Thus, $u(A)A^* \sim A$ and $u(B)A^* \sim B$.

Suppose $A \succcurlyeq B$. By transitivity of $\succcurlyeq$, $u(A)A^* \sim A \succcurlyeq B \sim u(B)A^*$ implies $u(A)A^* \succcurlyeq u(B)A^*$.

If $u(B) > u(A)$, then $u(B)A^*$ strongly dominates $u(A)A^*$. Since, $\succcurlyeq$ is weakly dominant, we get $u(B)A^* \succ u(A)A^*$ leading to a contradiction.

Thus, we get $u(A) \geq u(B)$, i.e., $A \succcurlyeq B$ implies $u(A) \geq u(B)$.

Now suppose, $u(A) \geq u(B)$.

If $u(A) = u(B)$, then $u(A)A^* = u(B)A^*$, so that $A \sim u(A)A^* = u(B)B^* \sim B$ combined with reflexivity and transitivity of $\succcurlyeq$ implies $A \succcurlyeq B$.

If $u(A) > u(B)$, then $u(A)A^*$ strongly dominates $u(B)A^*$. Since, $\succcurlyeq$ is weakly dominant, we get $u(A)A^* \succ u(B)A^*$.

Thus, $A \sim u(A)A^* \succ u(B)A^* \sim B$.

By transitivity of $\succcurlyeq$, we get $A \succcurlyeq B$.

Hence, $u(A) \geq u(B)$ implies $A \succcurlyeq B$.

Thus, $A \succcurlyeq B$ <u>if and only if</u> $u(A) \geq u(B)$. Q.E.D.